\documentclass{ieeeaccess}
\usepackage{cite}
\usepackage{amsmath,amssymb,amsfonts}
\usepackage{algorithmic}
\usepackage{graphicx}
\usepackage{textcomp}
\usepackage{caption}
\def\BibTeX{{\rm B\kern-.05em{\sc i\kern-.025em b}\kern-.08em
    T\kern-.1667em\lower.7ex\hbox{E}\kern-.125emX}}
\begin{document}


\title{Qubit Sensing: A New Attack Model for Multi-programming Quantum Computing}
\author{\uppercase{Abdullah Ash Saki}\authorrefmark{1}, \IEEEmembership{Student Member, IEEE}, and
\uppercase{Swaroop Ghosh\authorrefmark{2}, 
\IEEEmembership{Senior Member, IEEE}}}
\address[1, 2]{Pennsylvania State University, University Park, PA 16803 USA (email: axs1251@psu.edu)}
\tfootnote{This work is supported by National Science Foundation (NSF) (OIA-2040667, CNS-1722557, CCF-1718474, DGE-1723687 and 1821766) and seed-grants from Penn State ICDS and Huck Institute of Life Sciences for support.}


\corresp{Corresponding author: Abdullah Ash Saki (email: axs1251@psu.edu).}

\begin{abstract}
Noisy quantum computers suffer from readout or measurement error. It is a classical bit-flip error due to which state ``1'' is read out as ``0'' and vice-versa. The probability of readout error shows a state dependence i.e., flipping probability of state ``1'' may differ from flipping probability of state ``0''. Moreover, the probability shows correlation across qubits. These state-dependent and correlated error probability introduces a signature of victim outputs on adversary output when two programs are run simultaneously on a same quantum computer. This can be exploited to sense victim output which may contain sensitive information. In this paper, we systematically show that such readout error dependent signatures exist and that an adversary can use such signature to infer a user output. We experimentally demonstrate the attack (inference) on $3$ public IBM quantum computers. Using Jensen-Shannon Distance (JSD) a measure for statistical inference, we show that our approach identifies victim output with an accuracy of $96\%$ on real hardware. We also present randomized output flipping as a light-weight yet effective countermeasure to thwart such information leakage attack. Our analysis show the countermeasure incurs a minor penalty of $0.05\%$ in terms of fidelity.
\end{abstract}

\begin{keywords}
Quantum Computer, Readout error, Measurement error, Qubit, Sensing, Security, Privacy.
\end{keywords}


\maketitle
\pagestyle{plain}
\section{Introduction}
\label{sec:introduction}
\PARstart{T}{he} number of quantum bits (qubits) in Quantum Computers (QC) are increasing with IBM projecting 1000+ qubits device by 2023~\cite{ibm1000q}. Typically, $\approx50$ qubits are needed for quantum advantage~\cite{google-supremacy} i.e., to solve classically intractable problems. The current direction is to exploit the noisy qubits using approximate hybrid quantum-classical algorithms instead of using them for fault tolerance. Under such scenarios, large number of qubit might remain unused if a single program is executed in the hardware, degrading the hardware utilization and worsening the queue size. This gives rise to multi-programming~\cite{multiprogramming} environment where multiple users can run their workloads in parallel (on the same QC but different qubits). Multi-programming increases device utilization and financial return of QC vendors.  
\begin{figure}
    \centering
    \includegraphics[width=3.2in]{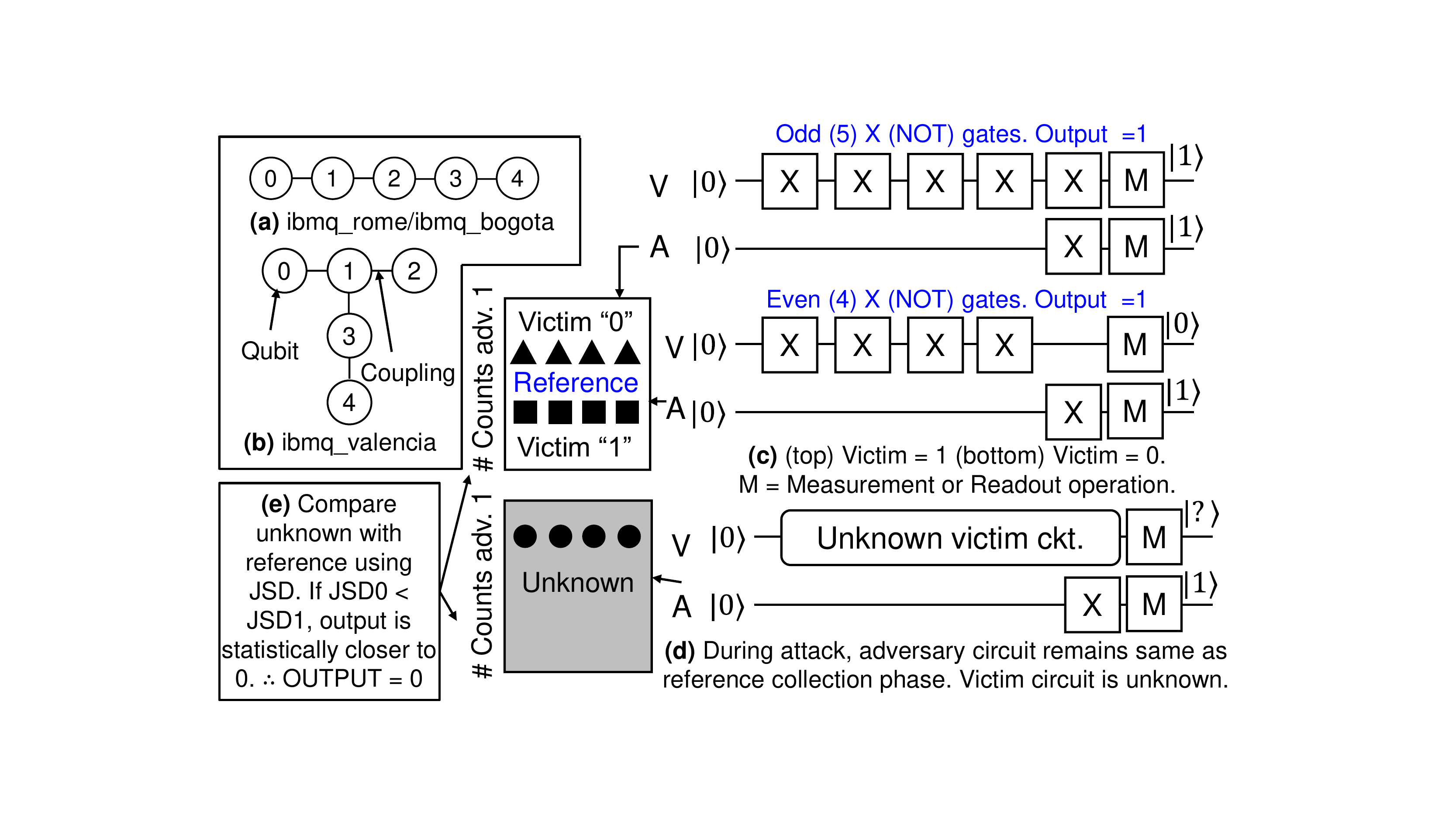}
    \caption{(a) and (b) Coupling graph of 3 IBM quantum computers. The circles are the qubits and the edge between two circles is the coupling. (c) Circuits for collecting reference readout signatures. We collect signatures for a victim in ``1'' and ``0'' states. The initial state of a qubit is ``0''. We choose odd (5 stages) and even (4 stages) number of X gates to prepare it to ``1'' and ``0'' respectively. (d) Circuits during an attack, and (e) The unknown signature from attack is compared with reference signatures to infer victim output.}
    \label{fig:overview}
\end{figure}

Although multi-programming is enticing, readout error in QC and its correlation with other qubit states in the device present a security vulnerability.
Readout error is a classical bit-flip error (e.g., output ``1'' is read out as ``0'' and vice-versa). The probability of the error is output state-dependent i.e., output 1 may flip more than output 0 and vice-versa. Moreover, the readout error probability of a qubit is correlated with states of other qubits in the system. Therefore, an adversary can exploit this correlation and infer the victim’s state by reading his/her qubit. The victim can be running algorithms with sensitive outputs (e.g., financial portfolio optimization) and one of the several objectives of an adversary is to steal such sensitive data for financial gain. 

\emph{To the best of our knowledge, this is the first work that experimentally demonstrates the vulnerability in multi-programming in QC and proposes a countermeasure}. This work will preserve the privacy of future large-scale QCs in a multi-programming environment. 

\textbf{Paper contributions:} We,
\begin{itemize}
    \item Collect readout signatures from multiple real hardware from IBM,
    \item Show correlation-based analysis to identify the victim’s computation result, and
    \item Present output obfuscation with randomized inversion as defense with overhead analysis.
\end{itemize}

The rest of the paper is organized as follows: In Section~\ref{sec:attack-method}, we present the attack model, method of reference signature collection and classification of victim output using statistical distance. 
In Section~\ref{sec:ccountermeasure}, we discuss a countermeasure and associated overhead. Finally, we draw conclusion in Section~\ref{sec:conclusion}.

\section{Proposed Attack Methodology}\label{sec:attack-method}
In the section, we present the attack model, methodology of reference signature collection, and classification of victim output using statistical method.

\subsection{Attack Model}
The victim and the adversary run their circuits on the same QC but different qubits. The adversary has access to reference signatures collected through prior experiments (more details in Section~\ref{sec:ref-sig-collection}) that shows distinct behavior for victim “0” vs. “1” state. During the attack, the adversary can only read his/her qubit. He/she then computes the statistical distance (Jensen-Shannon Distance) between the unknown victim data and the reference signatures.
\begin{figure}
    \centering
    \includegraphics[width=3.2in]{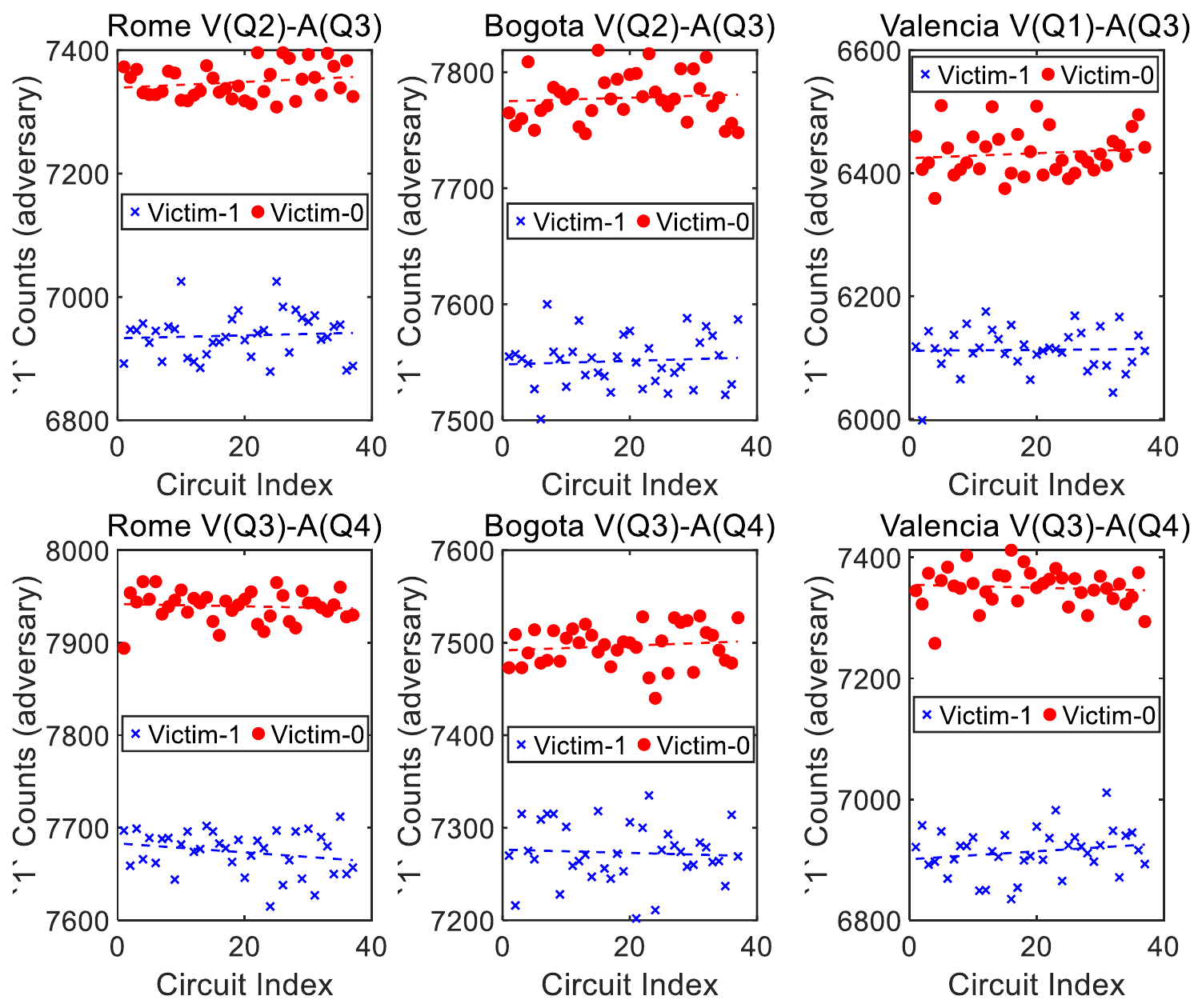}
    \caption{Experimental signatures for victim states 1 \& 0 from 3 IBM devices. Qubits on each device show a distinct signature for 1 \& 0.}
    \label{fig:2}
\end{figure}

\subsection{Reference Signature Collection}\label{sec:ref-sig-collection}
First, we establish that reading out 1 vs. 0 on a victim qubit has a distinct signature on the adversary qubit by running experiments on 3 different IBM quantum computers namely, ibmq\_rome, ibmq\_bogota, and ibmq\_valencia (Figure~\ref{fig:overview}a \& b). The circuits to collect the signatures are shown in Figure~\ref{fig:overview}c (victim qubit 1) and \ref{fig:overview}d (victim qubit 0). The victim program runs on one of the qubits and the adversary senses the outcome from another qubit. One circuit contains an odd while the other contains an even number of X (NOT) gates which puts the victim to ``1'' and ``0'' states, respectively. In both cases, the adversary qubit contains a single X-gate that puts it to “1”. The X gate on adversary qubit is applied at the last time-step to minimize the effect of qubit relaxation. 

As an example, consider the adversary qubit is QA and the victim qubit is QV. We set QA to 1 and QV to 0 and run it for 8192 trials (also known as shots). The ideal outcome should be 8192 number of 1s. However, the real outcome will be a mixture of 0s and 1s partly due to state-dependent readout error (the count of 1 in adversary qubit depends on the state of itself and the state of victim’s qubit). Suppose the output distribution of QA is V0. Next, we repeat the routine with QA at 1 (as before) and QV at 1. Suppose the output distribution of QA is V1. V0 and V1 will be the reference signatures/distributions for the pair QA-QV. For the same adversary-victim qubit pair, we collect multiple ($37 \times 2$) such reference signatures for each pair to minimize statistical errors during the inferencing phase (IBM devices allow up to 75 circuits in a single queue. Therefore, a maximum of 37 for each victim “1” and “0” reference circuits could be run at a time). The collected signatures from various victim-adversary pairs on 3 IBM devices (Figure~\ref{fig:2}) show that adversary counts vary between victim “0” and “1”. We use this asymmetry in adversary counts to sense the victim’s outcome.

\begin{figure}
    \centering
    \includegraphics[width=3.2in]{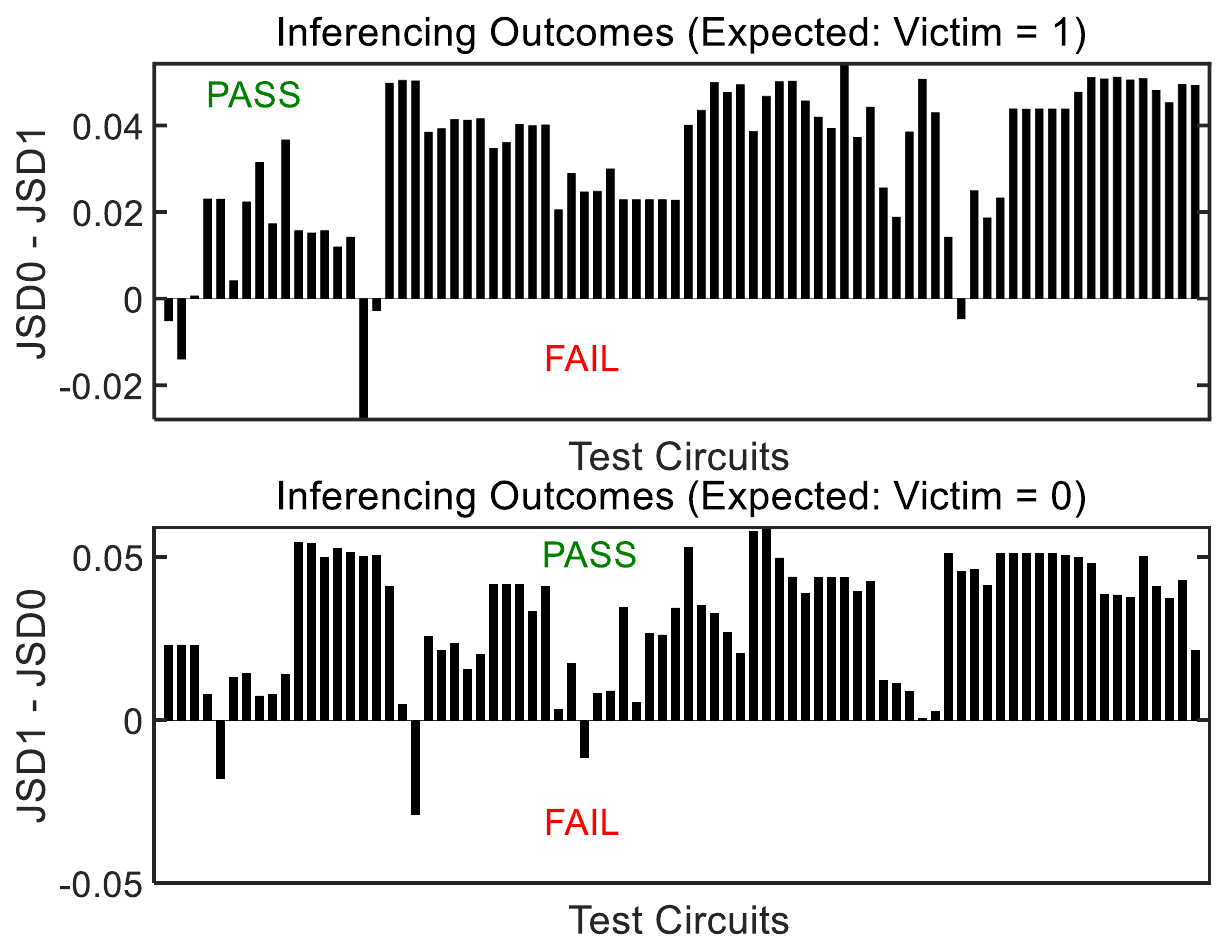}
    \caption{Experimental inferencing results for 3 devices. JSD0 = Jensen-Shannon-Distance between reference data for victim-0 and test-data and similar for JSD1. If JSD1 < JSD0, the test data is closer to reference-1 and it will be classified as 1 and vice versa. Among 200 test circuits, it correctly classifies victim state 192 times.}
    \label{fig:3}
\end{figure}

\subsection{Classifying Victim Output}\label{sec:classify-victim}
\subsubsection{Adjacent qubit}
Next, we use the reference distributions to classify unknown data using Jensen-Shannon Distance (JSD) ~\cite{jsd} as the statistical metric which is a well known distance measure in quantum computing to compare two distributions. In a real scenario, an adversary will collect a signature from his/her qubit (while the victim output is initially unknown). Say, the distribution is $V_{unknown}$. The adversary will compute two JSD values: JSD0 between \{V0 and $V_{unknown}$\} and JSD1 between \{V1 and $V_{unknown}$\}. If JSD0 > JSD1, then the unknown distribution is closer to V1 meaning the victim qubit is 1 and vice versa. 

We demonstrate the efficacy of the method by running several random circuits on the victim qubit and detecting outputs correctly. We choose randomized benchmarking (RB)~\cite{RB} circuits of varied depth (1-10) as our random circuits with state preparation. The state preparation block is either a X-gate (prepares qubit in 1) or I-gate (keeps the qubit in initial 0 state). RB circuit starts from an initial state and consists of m random gates. The first (m-1) gates put the qubit in an arbitrary state and the m$^{th}$ (last) gate inverts the arbitrary state and puts the qubit in the initial state. 
We use a single X gate to prepare the initial state “1” and an Identity (I) gate to prepare “0” at the start of the RB sequence. 
In all cases, the adversarial sensing circuit consists of a single X gate. We generate the RB circuits for victim qubit using Qiskit~\cite{qiskit} and test 200 different circuits on 3 devices. The victim output is correctly classified 192 times (accuracy 96\%). Figure~\ref{fig:3} shows $\Delta$JSD values for a sub-set (due to clarity) of the 200 test circuits with majority passing.

\subsubsection{Non-adjacent sensing}
\begin{figure}
    \centering
    \includegraphics[width=3.2in]{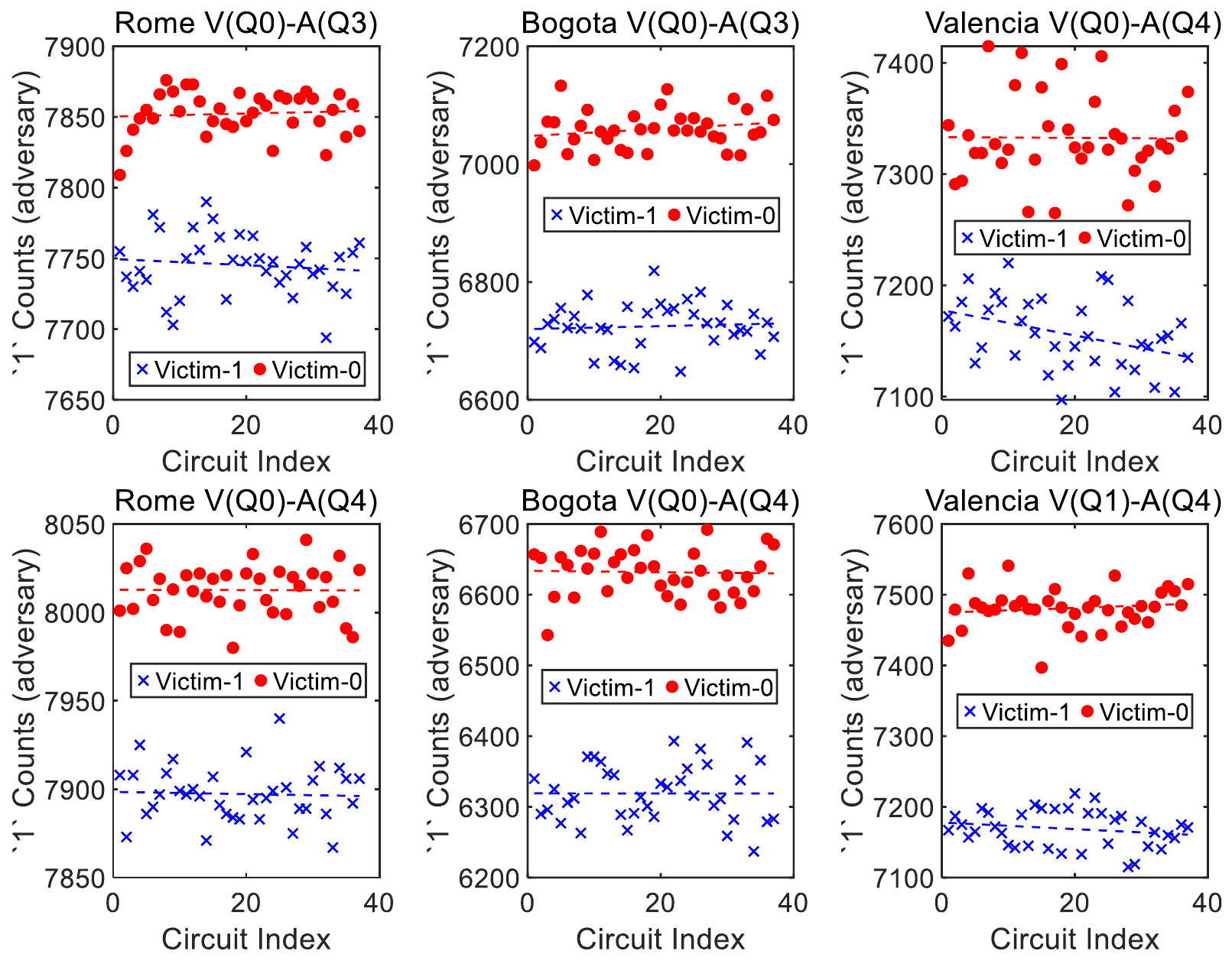}
    \caption{Readout signatures when victim and adversary qubits are non-adjacent. Even if qubits are far (e.g., Q0 and Q4 in ibmq\_rome), they show separate signatures for the victim being 0 and 1. Thus, proposed sensing can work even-if qubits are non-adjacent.}
    \label{fig:4}
\end{figure}

We collect data from many non-adjacent qubit pairs to show that the victim and adversary qubits need not be adjacent for the sensing. Figure~\ref{fig:4} shows that the physically distant qubits e.g., Q0 and Q4 in ibmq\_rome show distinct signatures.

\subsection{Sensing multiple victim qubits using a single adversary qubit}
\begin{figure}
    \centering
    \includegraphics[width=3.2in]{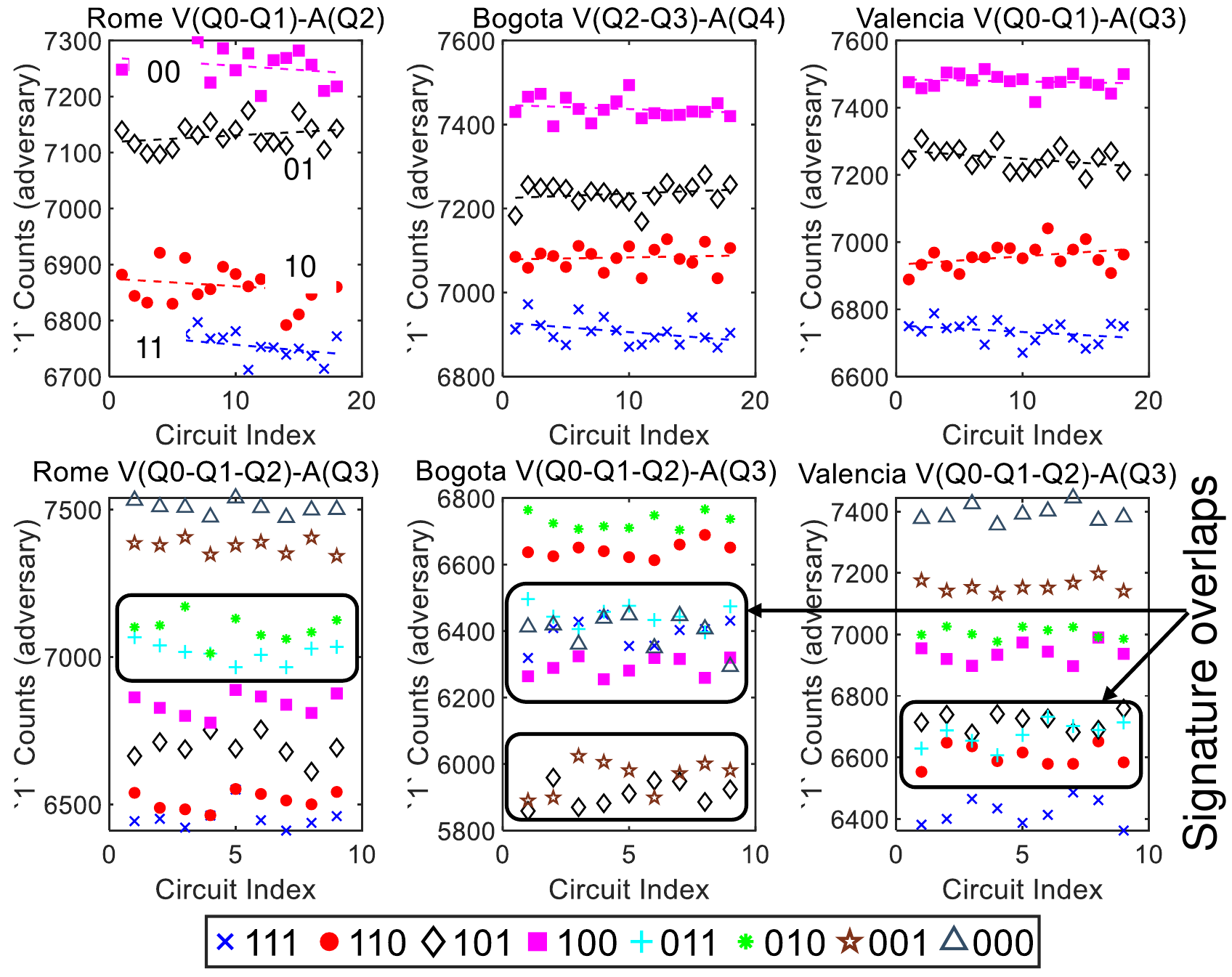}
    \caption{(Top panel) Sensing 2 victim qubits using 1 adversary qubit. The plots show 4 possible victim states (00, 01, 10, and 11) have a distinct signature. (Bottom panel) Sensing 3 victim qubits using 1 adversary qubit. Several signatures overlap with one another. Thus, a maximum of 2 qubits can be sensed using 1 adversary qubit.}
    \label{fig:5}
\end{figure}

Since qubits are connected to each other, it is possible to sense more than one victim qubit by the adversarial qubit. We extend our scheme to sense to validate this idea. First, we collect a signature for 2 victim qubit (+ 1 adversary qubit). There could be 4 possible outputs for 2 victim qubits (00, 01, 10, and 11). Adversary counts for these 4 outputs are separate as shown in Figure~\ref{fig:5}. It validates that by using a single qubit we can sense the output of multiple-victim qubits. 

To check the extent of sensing, we keep increasing the number of victim qubits to 3, 4, etc. Figure~\ref{fig:5} (bottom panel) shows the signature for cases with 3 victim qubits. In these cases, the signatures get closer and several overlaps with each-other rendering the sensing using only one qubit infeasible. Thus, we conclude that a single adversary qubit can sense only up to 2 qubits.

\section{Countermeasure}\label{sec:ccountermeasure}
In this section, we discuss a countermeasure to thwart the qubit sensing attack and discuss the overhead.
\subsection{Basic idea}
The adversary relies on the correlation between the collected and the reference data to infer the output (without knowledge of the victim’s original circuit and the output). The victim can randomly invert one or more qubit in his/her final output by inserting a NOT gate(s) (Figure~\ref{fig:6}a). Consider, the actual output was supposed to be 0. The victim adds the X-gate and reads it out as 1 (as the victim knows about the added X-gate, he/she can post-process the data and invert it back to 0). The adversary will sense it as 1 and consider it as the correct output which is wrong. Without knowing which qubit(s) victim has inverted, the adversary cannot use the signature to make an educated decision. He/she can then only resort to random guess. For example, for 1 victim qubit ($2^1 = 2$ possible outputs), random guess accuracy is 50\% (significant drop from 96\% above). For 2 victim qubits, it drops to 25\% and so on. To summarize, random inversion of qubit(s) can impose significant reverse-engineering effort on the adversary’s statistical approach to infer.
\subsection{Overhead analysis}

\begin{figure}
    \centering
    \includegraphics[width=3.2in]{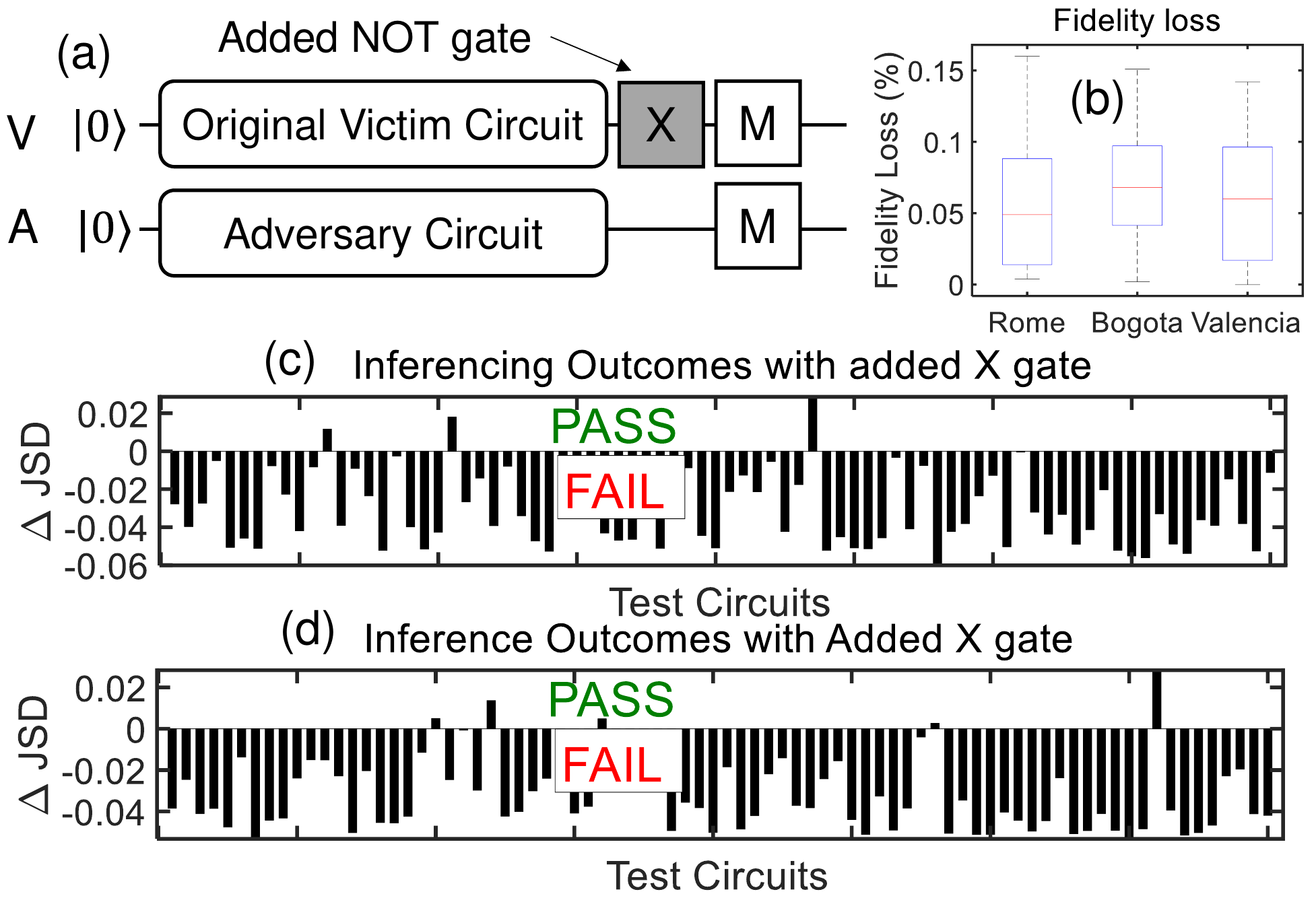}
    \caption{(a) Schematic of the victim circuit with additional X-gate to counteract sensing through adversary qubit, (b) Fidelity loss due to the added X-gate. The fidelity loss is about 0.05\% on average, (c) \& (d) The extra invert operation causes wrong inferencing.}
    \label{fig:6}
\end{figure}

Adding an X-gate adds gate-error in the circuit and increases its run-time. Both hurt the program fidelity. However, single-qubit gates have a lower error-rate $(10^{-4}-10^{-3})$ and a faster run-time ($\approx$35-50 ns). Therefore, the loss of fidelity is usually negligible. We define the fidelity as \{correct output count/total number of trials\}. We compute the fidelities with the added X-gate and without the X-gate. For test circuits, the average loss in fidelity is $\approx0.05\%$ (Figure~\ref{fig:6}b) from experiments on 3 IBM devices. Figure~\ref{fig:6}c \& d show that with the added X-gate the inferencing at the adversary end becomes wrong.

\section{Conclusion}\label{sec:conclusion}
Multi-programming computing model is attractive for future large qubit systems to improve the hardware utilization. We show that an adversary can sense another user’s output in such multi-programming QCs presenting privacy risk. We experimentally demonstrate the attack with 96\% accuracy and present randomized qubit flipping as a low-overhead circuit-level countermeasure.
\bibliographystyle{IEEEtran}
\bibliography{IEEEabrv,ref}
\EOD

\end{document}